\documentclass[preprint,11pt]{elsarticle}

\usepackage{epsfig}
\usepackage{array,tabularx,epsfig,mathrsfs,graphicx,rotating}
\usepackage{ifthen}
\usepackage{amsfonts}
\usepackage{ragged2e}
\PassOptionsToPackage{hyphens}{url}
\usepackage[hyphens]{url}
\usepackage{hyperref}
\usepackage{listings}
\usepackage{epstopdf}
\usepackage{color}
\usepackage{float}
\usepackage{subfig}

\usepackage[normalem]{ulem} 
\usepackage{soul} 
\usepackage{amsmath,amssymb}

\let\originallesssim\lesssim
\let\originalgtrsim\gtrsim

\DeclareRobustCommand{\lesssim}{%
  \mathrel{\mathpalette\lowersim\originallesssim}%
}
\DeclareRobustCommand{\gtrsim}{%
  \mathrel{\mathpalette\lowersim\originalgtrsim}%
}

\makeatletter
\newcommand{\lowersim}[2]{%
  \sbox\z@{$#1<$}%
  \raisebox{-\dimexpr\height-\ht\z@}{$\m@th#1#2$}%
}
\makeatother

\hypersetup{
  colorlinks=true,
  linkcolor=blue,
  citecolor=blue,
  urlcolor=blue
}

\graphicspath{{figs/}}

\newcommand{\beq}{\begin{equation}}
\newcommand{\eeq}{\end{equation}}

\newcommand{\Jpsi}{\mathrm{J}/\psi}

\chardef\til=126

\journal{ANL-HEP-173427}
\date{Feb 17, 2022}

\begin{document}
\definecolor{mygreen}{rgb}{0,0.6,0} \definecolor{mygray}{rgb}{0.5,0.5,0.5} \definecolor{mymauve}{rgb}{0.58,0,0.82}

\lstset{ %
 backgroundcolor=\color{white},   
 basicstyle=\footnotesize,        
 breakatwhitespace=false,         
 breaklines=true,                 
 captionpos=b,                    
 commentstyle=\color{mygreen},    
 deletekeywords={...},            
 escapeinside={\%*}{*)},          
 extendedchars=true,              
 keepspaces=true,                 
 frame=tb,
 keywordstyle=\color{blue},       
 language=Python,                 
 otherkeywords={*,...},            
 rulecolor=\color{black},         
 showspaces=false,                
 showstringspaces=false,          
 showtabs=false,                  
 stepnumber=2,                    
 stringstyle=\color{mymauve},     
 tabsize=2,                        
 title=\lstname,                   
 numberstyle=\footnotesize,
 basicstyle=\small,
 basewidth={0.5em,0.5em}
}

\begin{frontmatter}

\title{
Some aspects of impact of the Electron Ion Collider on particle physics at 
the Energy Frontier}

\author[]{S.V.~Chekanov}
\ead{chekanov@anl.gov}

\author[]{S.~Magill}
\ead{srm@anl.gov}

\address[add1]{
HEP Division, Argonne National Laboratory,
9700 S.~Cass Avenue,
Lemont, IL 60439, USA.
}

\begin{abstract}
This overview  describes several science cases at the Electron-Ion-Collider (EIC) experiment 
which are traditional to general particle physics. It has an emphasis on connections 
between future  measurements at the EIC and the 
physics topics explored at high-energy frontier colliders.
It covers several selected topics, such as parton density functions, multi-quarks states,
correlations of final-state particles, precision QCD measurements and forward jet physics.
We will discuss possible EIC measurements that can improve previous experimental results 
obtained at high-energy (HEP) experiments.
\end{abstract}


\end{frontmatter}

\section{Introduction}

The Electron-Ion-Collider  (EIC) \cite{Accardi:2012qut,eic1} is a proposed collider experiment to study the 
structure of building blocks of matter -  the nucleus and the nucleons (protons and neutrons). 
This future experiment will collide beams of 
spin-polarized electrons with intense beams of both polarized 
nucleons and unpolarized nuclei.
The EIC is  a flagship nuclear science facility,  however, 
many goals of this project are aligned with high-energy physics (HEP) 
that  also studies the structure of the nucleons at their deepest level. 
In particular, EIC can revolutionize our understanding of quantum chromodynamics, a theory that is important for almost all physics processes studied at the energy frontier.  

This contribution describes science cases at the EIC experiment 
which are traditional to general particle physics,
with a particular emphasis on the connections to the physics at the energy frontier studied in previous circular colliders, such as the Large Electron–Positron Collider (LEP),  the 
Hadron-Electron Ring Accelerator (HERA),  the Large Hadron Collider (LHC) and the Tevatron experiments at the Fermi National Accelerator Laboratory (Fermilab).
The paper  describes a number of possible measurements related to 
parton density functions (PDF), multi-quarks states, precision QCD measurements and forward jet physics.
In addition to rather traditional topics, such as perturbative QCD, hadronic jets and  PDFs,
a significant area of research  at the energy frontier is soft (i.e. non-perturbative) phenomena, and their interplay with the hard QCD. Therefore, this report also covers various not-well understood soft effects encountered in $ep$ and $pp$ collision experiments, that can be verified and further studied at the EIC experiment.

\section{Parton distributions functions}

The knowledge of parton density functions (PDF) has a crucial impact on future collider program of HEP. The EIC is expected to supersede the previous fixed-target deep-inelastic scattering  (DIS) experiments that currently dominate the constraints on high-$x$ PDFs. The EIC will help constraining PDF dependence of HEP observables at moderate and large $x$, including the Higgs and electroweak sectors like $M_W$ and $\sin^2 \theta_W$. The effect of measured PDF at EIC show substantial sensitivity to the total Higgs production cross section at the LHC \cite{Hobbs:2019sut}.

Recent re-analysis of HERA data \cite{Abt:2020mnr} with the focus on high-$x$ regions of the proton PDF indicated that the uncertainties on the current PDFs are underestimated. At the value of $x$ larger than 0.5 (and $525<Q^2<9500$~GeV$^2$), such uncertainties could be significantly larger than currently thought. The EIC experiment can clarify this question by providing high-precision data for $x>0.5$ in the region of $10<Q^2<1000$~GeV$^2$, i.e. with an overlap in the ($Q^2-x$) kinematic plane with the HERA data at low $Q^2$ studied in \cite{Abt:2020mnr}.

\section{Beauty and charm production}
\label{flavour}

\begin{figure}[htb]
\begin{center}
\subfloat[Charm cross section]{
\includegraphics[width=0.45\textwidth]{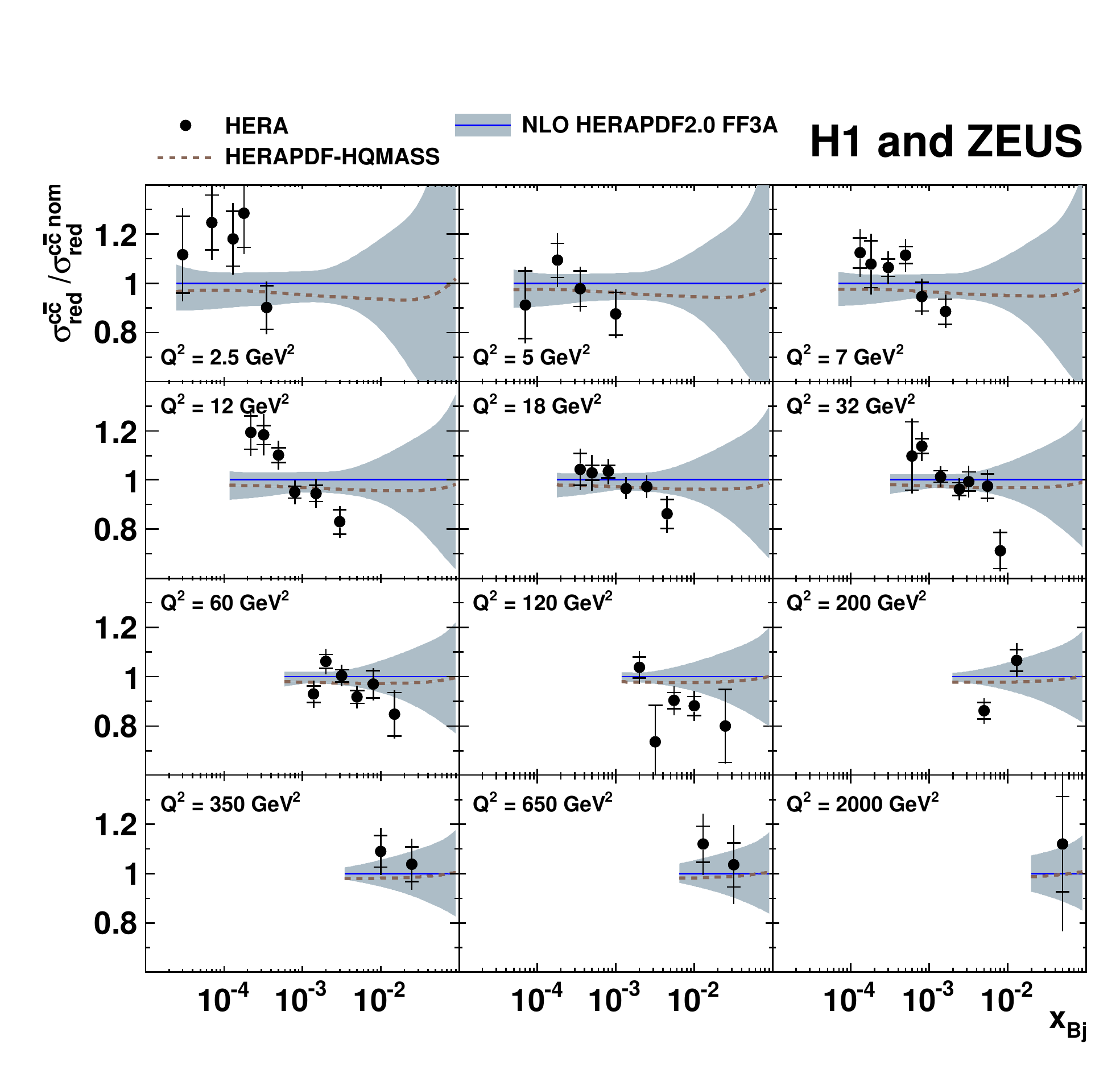}
}
\subfloat[Beauty cross sections]{
\includegraphics[width=0.45\textwidth]{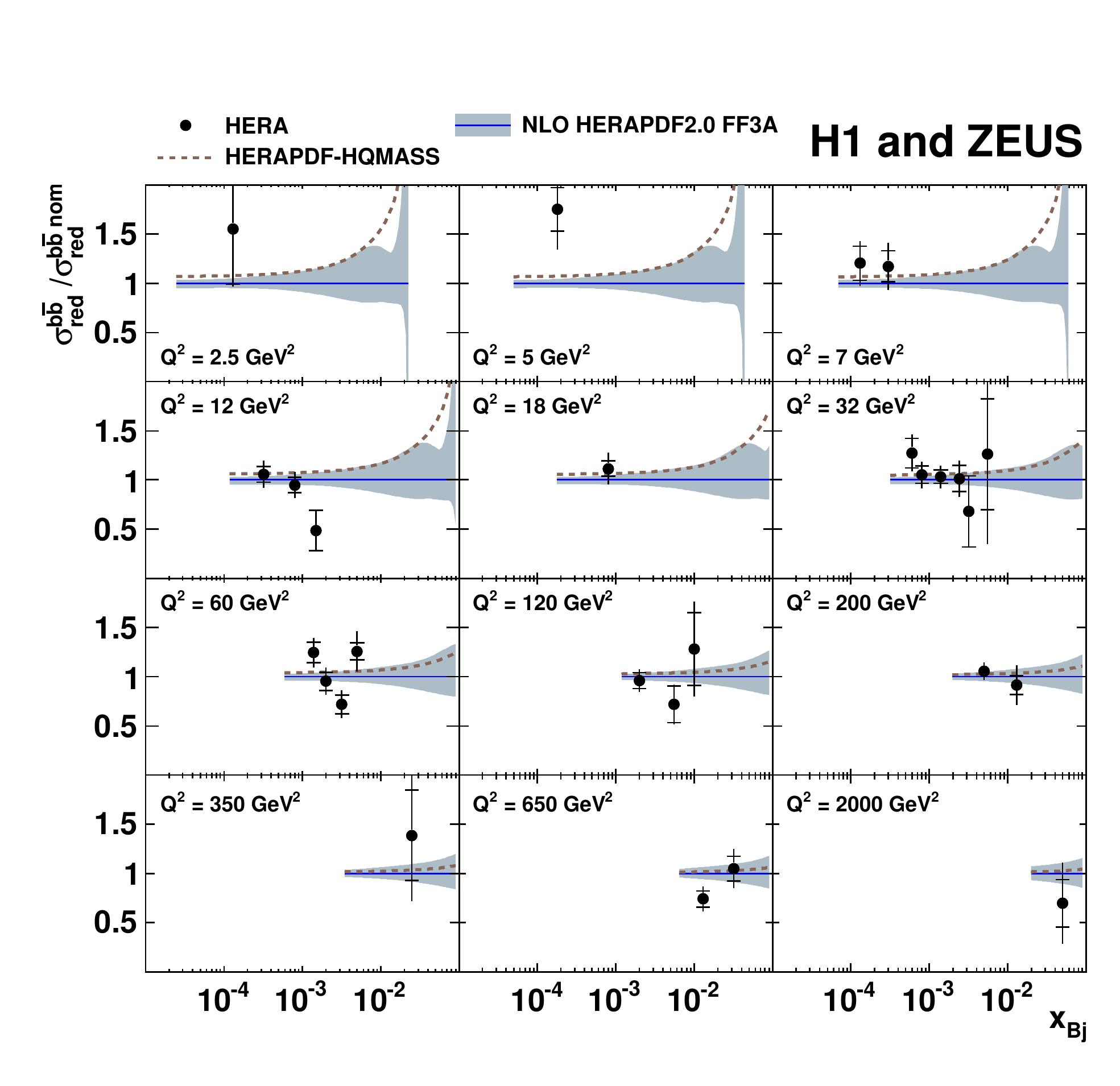}
}
\end{center}
\caption{
Ratio of reduced charm (top) and beauty (bottom) cross sections, 
as a function of $x$ for given values of $Q^2$, to the NLO FFNS 
predictions using HERAPDF-HQMASS (dash lines) with 
respect to the reference cross sections $\sigma$ compared, based on HERA-APDF2.0 FF3A (solid lines with uncertainty bands). From Ref.\cite{H1:2018flt}. }
\label{fig:d18-037f16}
\end{figure}

Measurements of cross sections in particle physics provide  an important testing ground for calculations of the QCD matrix elements and our understanding of PDFs.  The studies of  open  charm  and  beauty  production  in  neutral current  (NC)  deep  inelastic scattering (DIS) by H1 and ZEUS \cite{H1:2018flt} show the consistency of experimental data with theoretical calculations
performed at the  next-to-leading (NLO) and the approximate next-to-next-to-leading-order QCD predictions based on different flavor schemes. However, the data on  beauty production have significant experimental uncertainties that can be reduced at the EIC. 
Figure~\ref{fig:d18-037f16} illustrates
that the uncertainties on charm (beauty) cross sections at low $Q^2$ are at the level of 20\% (50\%). Data at high $x$ at $2.5<Q^2<60$~GeV$^2$ are not available. This is precisely where the theory has large theoretical uncertainties, and where the EIC can provide valuable
data for testing such predictions.

Even for the given level of uncertainties, a recent QCD analysis \cite{H1:2018flt} reveals some tensions at the level of $3~\sigma$ in describing simultaneously the inclusive and the heavy-flavour HERA DIS data. This can also be resolved at the EIC.

\begin{figure}[htb]
\begin{center}
\includegraphics[width=0.7\textwidth]{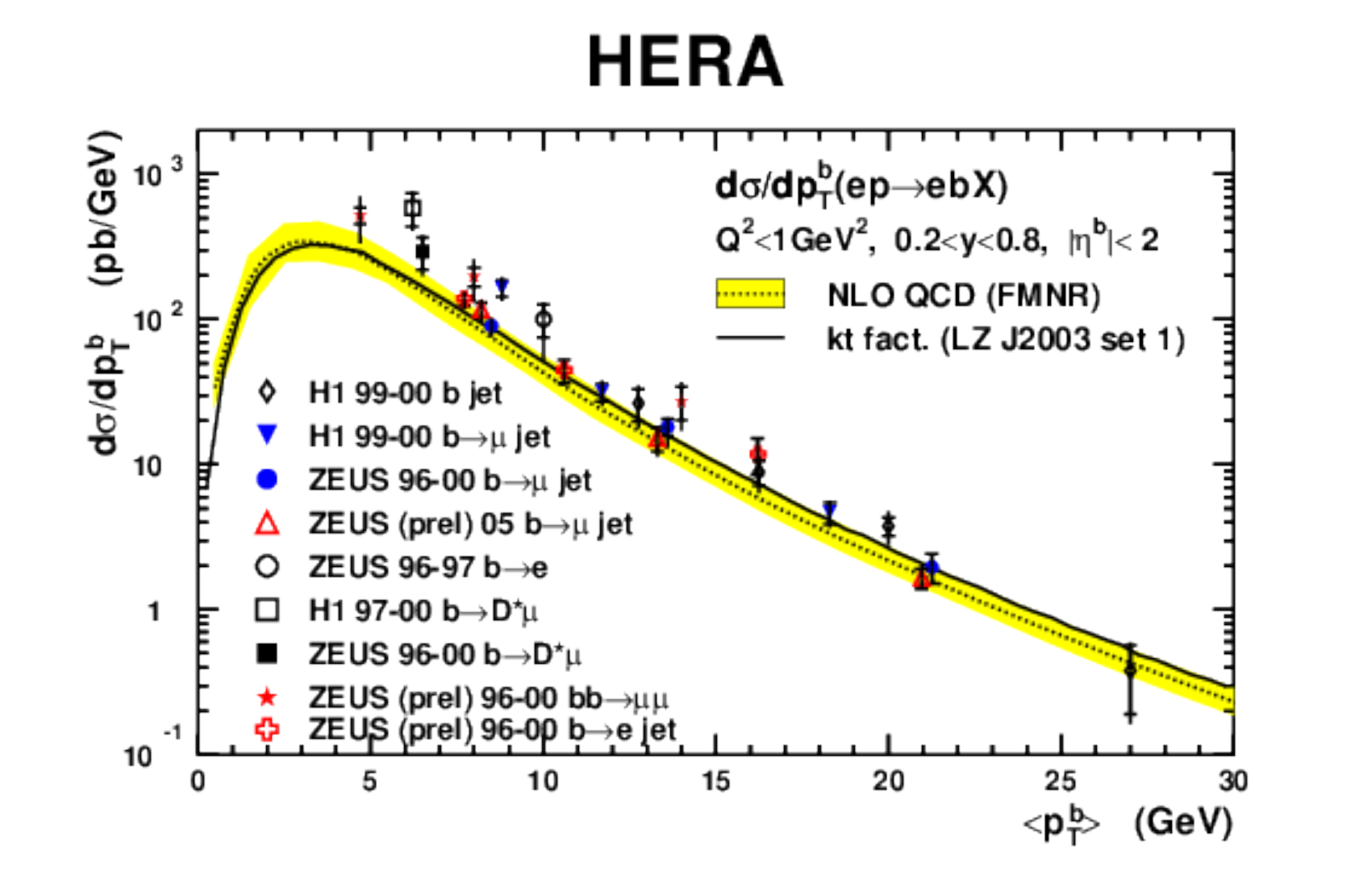}
\end{center}
\caption{
A summary of HERA data on beauty photoproduction cross-section
$d\sigma/dp_T^b$ compared to NLO QCD in the massive scheme and a
$k_T$ factorisation calculation (see \cite{Klein_2008} and referenced herein). }
\label{fig:bmesons}
\end{figure}

The EIC experiment can also provide new data on photoproduction of $b$-mesons.
Fig.~\ref{fig:bmesons} shows the $d\sigma/dp_T^b$ cross section
for  $b$-photoproduction compared to NLO QCD in the massive-quark scheme.
The measurements are somewhat above the theory at low transverse momenta, 
but they are still consistent with the predictions.  
No data in the region below $5$~GeV are available, i.e. near the 
bottom-quark mass of about 4.2~GeV.
This is the kinematic region where theoretical calculations should treat the heavy quark as a massive quark, and should include careful estimation of high-order QCD corrections.
The EIC data can fill this gap in terms of low-$p_T$ data.

\section{Fragmentation}

Fragmentation functions parameterize the transfer of the quark’s  energy to a given  hadron.  In this section we will discuss several effects that have been observed in the past, and which can be
further investigated at the EIC. 

\subsubsection{Charm fragmentation functions}

\begin{figure}[htb]
\begin{center}
\includegraphics[width=0.7\textwidth]{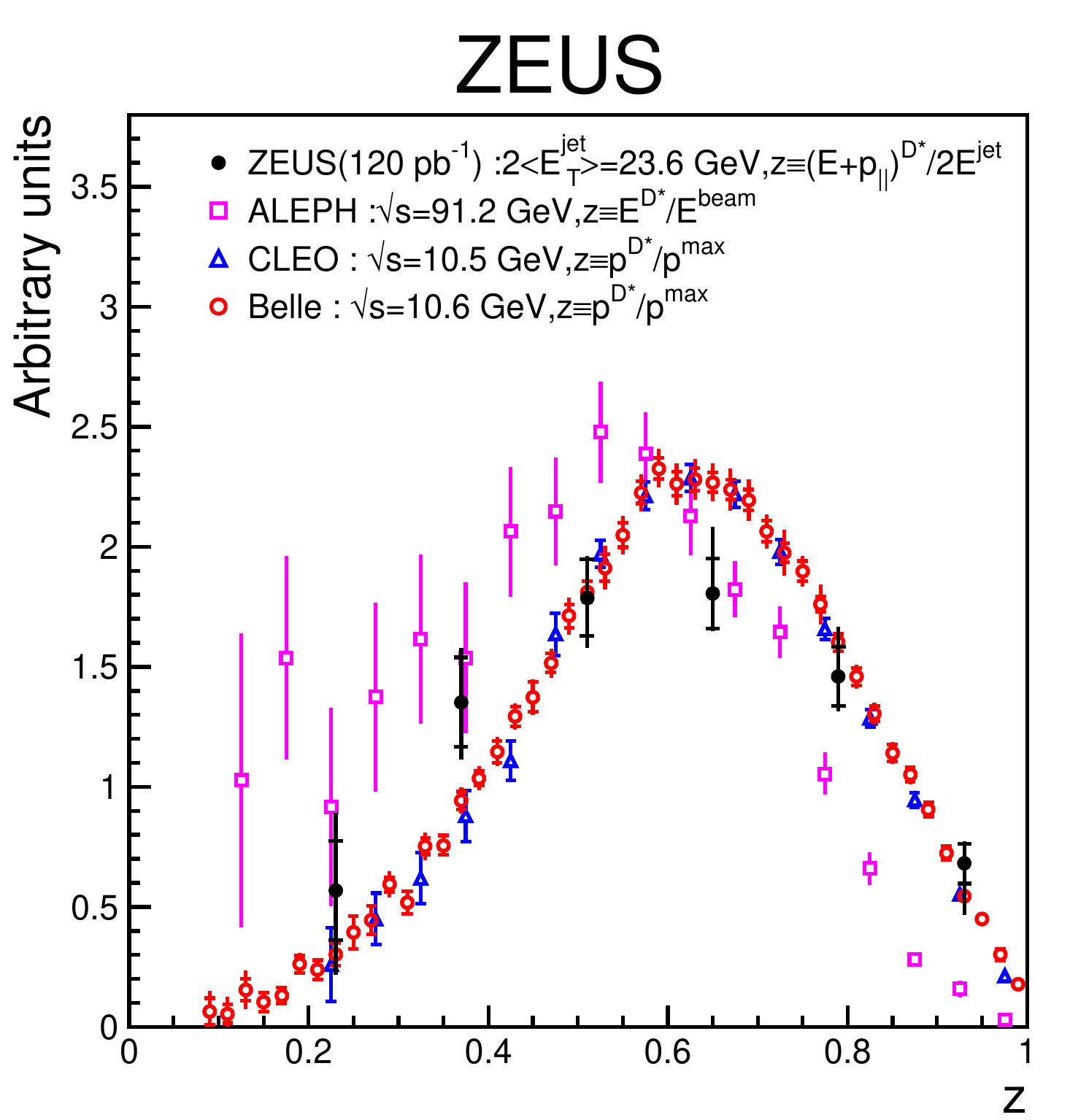}
\end{center}
\caption{$D^*$ fragmentation function measured  \cite{Chekanov:2008ur} at HERA compared to
measurements of $e^+e^-$ experiments. 
The fragmentation function is measured versus the ratio of the momentum
of the $D^*$ meson and the maximum momentum at the relevant beam energy.}
\label{fig:frag_zeus}
\end{figure}

Measurements  of fragmentation  functions and their comparison  with  $e^+e^-$ experiments  provide  a measure  of the  universality  of charm  fragmentation, and can further constrain its form.
At HERA, charm fragmentation functions measured using the $D^*$ production are significantly 
limited by available statistics \cite{Chekanov:2008ur}. 
Figure~\ref{fig:frag_zeus} shows  the $D^*$ fragmentation function versus the ratio of the momentum
of the $D^*$ meson and the maximum attainable momentum at the relevant beam energy.
Such results indicate that the precision in the $ep$ collision 
measurements is a factor 5 smaller than that in   $e^+e^-$ experiments. The EIC is expected to improve the precision of such measurements.

It is interesting to note that the normalized differential cross sections in 
the kinematic-threshold region cannot be well described \cite{Aaron:2008ac} by the NLO QCD calculations in the fixed-flavor number scheme using the independent fragmentation of charm quarks to $D^*{\pm}$. This indicates the need for realistic QCD calculations in all
kinematic regions of the  $D^*{\pm}$ measurements.

\subsubsection{Compound states}
Production of baryonic states in fragmentation is typically described by various phenomenological models.  Beyond multi-quark states, the production of  nucleus of  atoms in collider experiments, such as deuterons, which are composed of one proton and one neutron, cannot be calculated from first principles. According to physical cosmology, deuterium is expected to be produced in the first minutes that followed the Big Bang. Studies of deuterons and more complex states  in colliding experiments may shed light on the dynamic mechanism of the production of deuterium in the early universe.

Coalescence models are typically used to explain the production of deuterons after the hadronization step in particle colliding experiments, but they can only be used for a handy parameterization of observations, rather than for quantitative predictions. First observation of (anti)deuterons in DIS events of $ep$ collisions was reported at HERA \cite{Chekanov:2007mv}  with a puzzling observation of an excess of the deuteron rate over that of antideuterons  in the central fragmentation region where the contribution from incoming protons should be negligible. Another interesting observation is related to the measured parameter $B_2$,  reflecting the spatial size of the fragmentation region emitting the particles. In DIS, it is significantly larger than for heavy-ion experiments, but lower than for $pp$ and $pA$. 

The EIC experiment can  improve the existing measurements  due to the advanced particle identification in the regions of large $p_T/M$ and due to the large statistics of event samples.

\section{Short and long-range correlations}

Historically, short- and long-range correlations were an active area of studies of various collective effects in HEP.  

Short-range correlations of final-state particles are sensitive to interdependence between particles as the phase-space separations between them decrease. 
The correlations are related to fluctuations of particle densities in small rapidity windows, and are often
interpreted as a manifestation of ``intermittency'' in particle density 
that exhibits ``self-similarity'' 
with respect to the size of the phase-space volume \cite{intermit}.
Such local fluctuations have been intensively studied in $e^+e^-$, $ep$, $pp$, $AA$, $\mu p$, and $hA$ collisions 
(for a review see \cite{kittel}). Their strength  depends on  parton-evolution dynamics, 
Bose-Einstein correlations (to be discussed later),  resonance decays and other effects. Generally, short-range correlations and fluctuations can be reproduced by  QCD-based Monte Carlo models with various degrees of success \cite{kittel}.
An advanced two-track resolution (i.e. the ability to distinguish two nearby track) combined with 
a large event sample  can improve such measurements at small phase-space regions (in rapidity, angular variable etc.) that were not accessible by the previous experiments.

Long-range correlations in the Breit frame \cite{breitframe} of $ep$ collisions (the so-called current-target correlations), to the first-order perturbative QCD, can  be used to extract the behavior of the boson-gluon fusion rates as a function of the Bjorken variable $x$ \cite{1999LR}.
Such rates, in turn, depend on the gluon density inside the proton.
The measurements do not require explicit 
reconstruction of jets, therefore,  they are well suited for relatively low $Q^2$ events to be collected at the EIC where
jet reconstruction has significant uncertainties.
The studies performed by the ZEUS collaboration \cite{2000ZEUSLR} show that the correlation are large and negative. They are thus very different from those measured in $e^+e^-$, where  small and positive forward-backward
correlations have been observed. Some popular Monte Carlo simulations  fail to describe 
the long-range correlations in the Breit frame of $ep$ collisions. 

The long-range correlations in the Breit frame can also be viewed  \cite{Akushevich:1999dk} as a tool for high-precision measurements  of the polarized gluon distribution function $\Delta G$. Multiplicity correlations between current and target regions in the Breit frame  of DIS events are particularly sensitive to $\Delta G$ for polarized beams. 
It is important to emphasize again that this approach is well 
suited for a low $Q^2$ and small transverse momenta of jets, i.e. in the regions where dijet reconstruction suffers from large hadronization corrections and various instrumental effects, i.e. in the EIC collision environment. 

Recently, a significant attention has been  directed toward long-range
(large pseudorapidity gap) particle correlations that can arise from collective phenomena in high-energy collisions.
For example, such correlations can originate from the formation of a hot, strongly interacting quark gluon plasma with nearly ideal hydrodynamic behavior. 
The long-range correlations were observed in $pA$ and $AA$ at  the Relativistic Heavy Ion Collider (RHIC) and the LHC.
Surprisingly, $pp$ collisions at the LHC also indicate this type of correlations (see \cite{collaboration2020studies} for the recent results).
However,  the long-range correlations are  absent in  DIS $ep$ collision events, as recently studied in two-particle azimuthal correlations at HERA \cite{Abt_2020}. Therefore, the kind of collective behavior recently observed at RHIC and the LHC energies in high-multiplicity hadronic collisions does not exist in $ep$ collisions.

The EIC experiment can contribute to a better understanding of this effect due to  the unique opportunity of measuring the long-range correlations in both $ep$ and $eA$ collision using the same instrumental environment and the same  kinematic ranges.
The collectivity effect has never been explored in $eA$ collisions.
A reduced systematic when comparing two different processes using the same instrumental apparatus may  shed light on the evolution of the multi-particle production mechanism.

\section{Bose-Einstein correlations}

Bose-Einstein (BE) corrections play a significant role 
in short-range correlations between identical particles.
The BE effect has been extensively studied 
in heavy-ion, $e^+e^-$, $pp$, $p\bar{p}$ and $ep$ collisions.
These correlations  are known to reflect  the  spatial  dimensions  of  the production source 
from which the identical bosons originate. 
The results \cite{Adloff:1997ea,Chekanov_2004} from the HERA experiments indicate that the production source radii in $ep$ DIS are similar to those in $e^+e^-$,
but disagree with the source size in 
relativistic heavy-ion collisions where the radii depend on the atomic number $A$. It has been 
also found that the source of identical particles has an elongated shape \cite{Chekanov_2004}, consistent
with the expectations of the Lund model \cite{Andersson:1983ia}. In $ep$ collisions, the 
average size of the production source is typically smaller than $0.8$~fm, but it
can be as large as $1$~fm  in the longitudinal direction. For $pp$ collisions, BE correlations have a larger radius (up to 3~fm)  \cite{Sirunyan:2017ies,Sirunyan_2020,Aad_2015} than for $ep$ and $pp$ collisions. The EIC experiment is expected to update such measurements with a better precision and reduced systematics.

\section{Instanton-induced processes}

Searches for instanton-induced processes in DIS was one of the  well-established studies  at HERA. Such studies can bring a better understanding of  tunnelling transitions between topologically different vacua.  Instanton processes are a direct  prediction of the Standard Model, but such processes
have not been experimentally observed so far despite significant efforts to find such processes in $ep$ and $pp$ collisions (see the recent  overviews \cite{Khoze_2020,Khoze_2021}).
 
Recently, the H1 collaboration has updated their searches  for instanton-induced events \cite{instH1} using HERA II data, setting  the 
upper limits on the cross section for instanton-induced processes between 1.5~pb  and 6~pb at 95\% confidence  level.
 
The reason why searches for such events in DIS may be more attractive than in  $pp$ collisions is because of a relatively clean, gluon reach environment compared to $pp$ collisions, and because of  several sophisticated theoretical predictions, implemented in Monte Carlo simulations, are available for DIS (see references in  \cite{instH1}). 
It was argued \cite{Khoze_2021} that, although the instanton cross sections are very large at hadron colliders, experimental sensitivity to instantons is limited due to the trigger criteria that is necessary to reduce the event rate. 
DIS events  also have independent kinematic scale, 
such as highly virtual momentum scale $Q$. This 
can  make searches for instantons easier at EIC 
compared to $pp$ collisions at the LHC.

Observations of instantons in particle collisions,  a direct consequence of QCD, 
would be a major breakthrough in modern physics. The EIC experiment, in terms of accumulated integrated luminosity, should have a sensitivity to instanton-induced events at least a factor 10 better compared to HERA  \cite{Khoze_2021}. 
In the case if no evidence for such events  will be found,
experimental limits will be set at the level of a few tens of femtobarn ($fb$), 
which may reshape the way we understand
the tunnelling transitions between topologically different vacua.

\section{Particle spectroscopy}
\label{partspec}

The contribution of the EIC to the particle spectroscopy is a special topic that needs to be discussed. This topic is so broad  that we will only illustrate one measurement. It has been previously extensively conducted
in $ep$ \cite{Chekanov:2008ad}, heavy-ion collisions \cite{Aaij:2020qfw} and two-photon collisions in $e^+e^-$ experiments \cite{ACCIARRI2001173}.

Studies of the rich mass spectra of $K_0^S K_0^S$ touch several important areas that have been forming the core of particle physics in the past decades, such as the production of $f_0$, $f_2$ mesons and their branching rations, formation of possible glueballs or tetraquarks and their contributions to the meson spectra (for a review see \cite{Klempt:2007cp}) and  heavy flavour physics (such as the studies of branching fraction of the decay of $B^0$  \cite{Aaij:2020qfw}). 

\begin{figure}[htb]
\begin{center}
\subfloat[$K^0_S K^0_S$ in $ep$ collision at HERA]{
\includegraphics[width=0.45\textwidth]{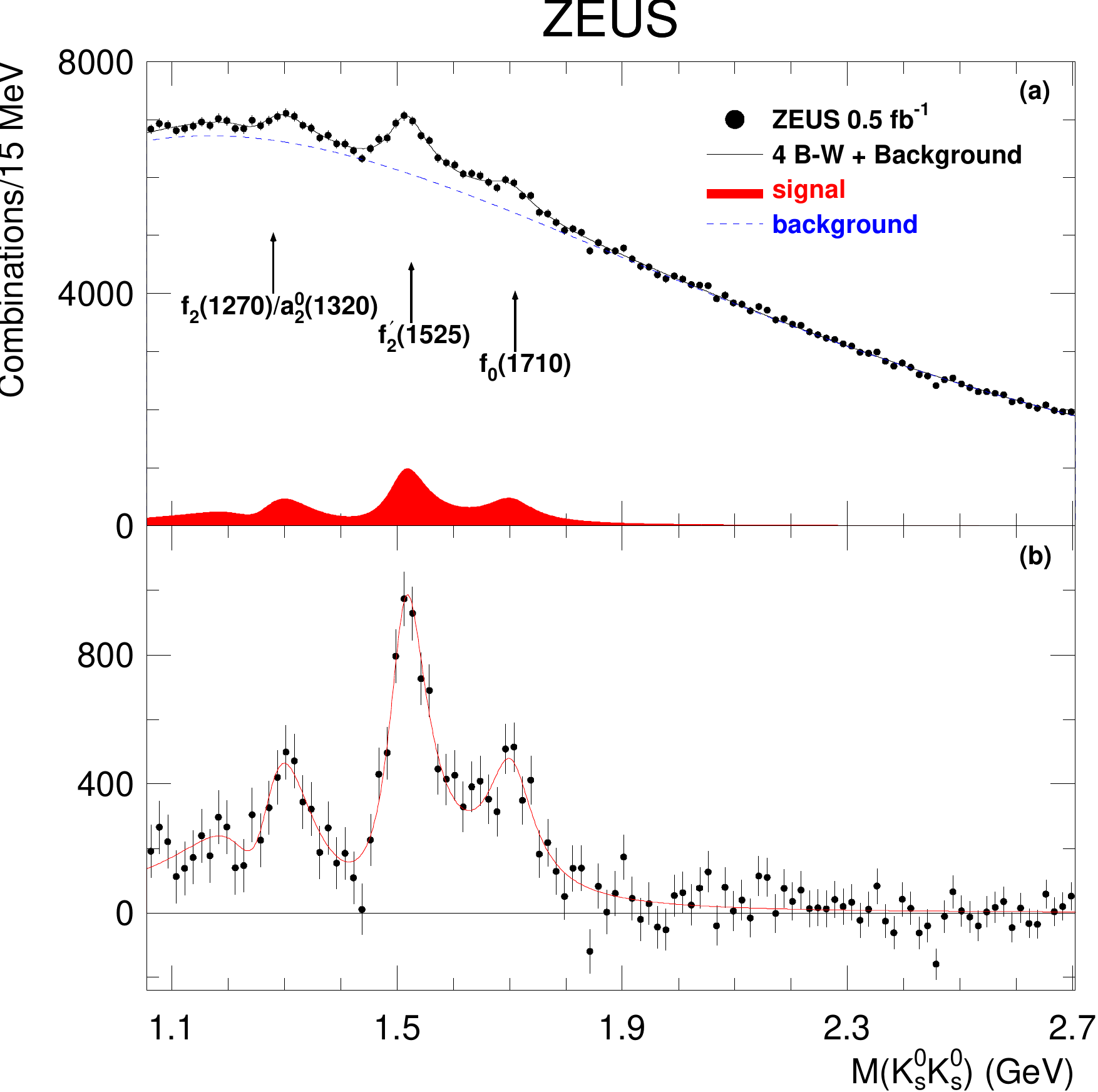}
}
\subfloat[$K^0_S K^0_S$ in $pp$ collision at the LHCb]{
\includegraphics[width=0.45\textwidth]{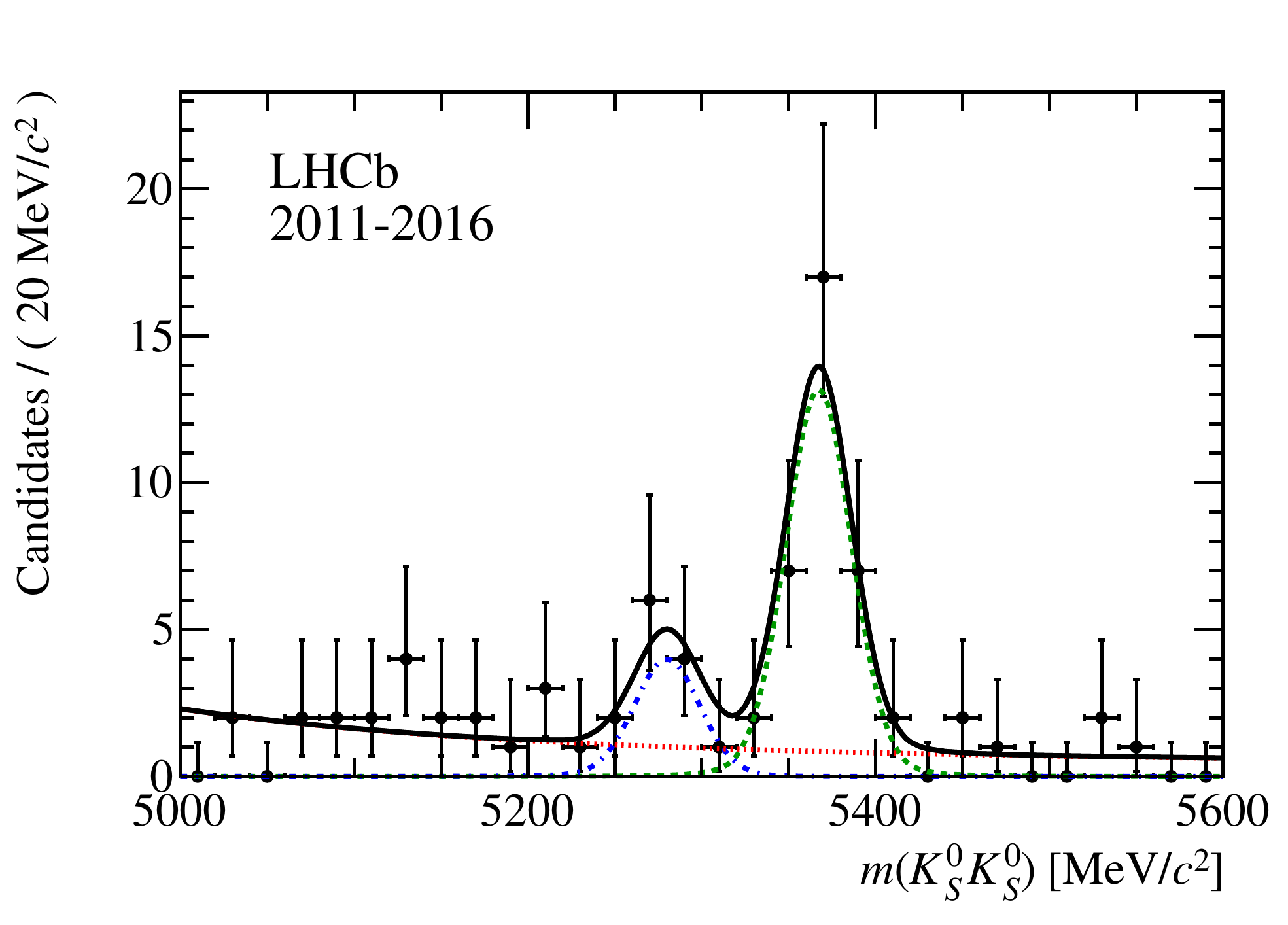}
}
\end{center}
\caption{
Invariant-mass distributions of $K^0_S K^0_S$ in $ep$ collision at ZEUS \cite{Chekanov:2008ad} and the LHCb \cite{Aaij:2020qfw}}
\label{fig:ksks}
\end{figure}

Figure~\ref{fig:ksks} illustrates typical invariant masses of the $K^0_S$ pairs in $ep$ and $pp$ collisions.
The measurements are severely limited by the available 
statistics  and, therefore, it can be improved at the EIC with 
advanced tracking systems and a high-efficient secondary vertex finding algorithm.

The $K^0_S K^0_S$ pairs provide  a clean sample of well-identified particle pairs that can also be used for searches of the CP violating processes in charmonium decays \cite{BAI20047},  such as $\Jpsi \to  K^0_S K^0_S$ or 
$\upsilon (2S) \to K^0_S K^0_S$. Similarly, the BaBar collaboration searched for the CP-violating
asymmetries for $B$ decays into charmless two-body final states \cite{Aubert:2001am}, i.e.  
$B^0 \to  K^0_S K^0_S$. The experimental upper limit on the branching fraction for such decays are typically of the order of
$10^{-6}$, and  is approaching the upper range of the current theoretical estimates.

The EIC can further improve these measurements. The expected number of $\Jpsi$ at the EIC
for 100 fb$^{-1}$ is in the range $(20-70)\times 10^6$ for  $y > 0.001$, depending on the CM energy planned for the EIC  This was estimated \cite{Sylvester} using a leading-order Monte Carlo simulation for all decay channels. This statistics is in direct competition with  BES data \cite{BAI20047}.

\section{Search for Lepton Flavour Violation}

Lepton flavour violation studies in the previous HERA experiments  \cite{Chekanov:2005au,Aaron:2011zz} focused on the processes $ep\rightarrow lX$, where
$l$ is a $\mu$ or $\tau$. The main feature  of such  events  is the presence of an  isolated $\mu/\tau$  with  high transverse  momentum, which is balanced by that of a jet in the transverse plane. 
An observation of such processes would  be  an indication 
for a signal  of  new physics  beyond  the  Standard Model. 
A typical model that explains such events would be a leptoquark (LQ) particle that
couples directly to leptons and quarks. Therefore, events like  $ep\rightarrow l\mu$,
can be explained by the exchange of leptoquarks. 

The HERA studies  indicate that the reconstructed leptoquark mass in the $\mu$ and $\tau$ channels agree well 
with the SM prediction (see Fig.~1 of \cite{Aaron:2011zz}).
However, the signal region contains only 10 events, therefore, 
such a comparison with the theory is statistically limited. The EIC experiment is expected to improve such measurements.



\section{Multiquark states}

The integrated luminosity expected at the EIC will be sufficient to probe particle production in significant  details, 
and can answer the questions related to the dynamics of multiquark state formation.  Composite particles can be either mesons (formed of quark–antiquark pairs) or baryons (formed of three quarks). Particles with a larger number of quarks are called pentaquarks. It should be noted that neither the quark model nor the QCD exclude the existence of such  non-conventional hadrons.

The LHCb experiment at the LHC reported several particles in the mass range of 4300 -- 4500~GeV  
decaying to a proton and a $\Jpsi$.  Such particles can be interpreted as  charmonium pentaquarks. More recently, the LHCb experiment made an observation \cite{Aaij:2020fnh} of a narrow structure around 6.9~GeV in the di-$\Jpsi$ invariant mass, matching the lineshape of a resonance. If it is confirmed, it  can be interpreted as a tetraquark state. The observation was obtained using 34,000 $\Jpsi$ pairs.

With a total cross section of about 10~nb at the EIC, the total number of produced $\Jpsi$ particles for 250~fb$^-1$ of integrated luminosity, the number of produced  $\Jpsi$ particles will be around 2.5 billion. Even after taking into account branching ratio and acceptance correction, the statistics of  reconstructed will be competitive to that of the  LHCb. 

An observation of  lighter pentaquaks with strange quarks can lead to a more complete picture of the QCD dynamics leading to such states. However, experimentally, searches for light pentaquarks in the LHC data is more challenging than searches  for heavy pentaquarks due to the complex collision environment at the LHC and due to high rates of light particles from fragmentation. 

Search for strange and light charm pentaquarks was a significant part of the scientific program at HERA.
An evidence for strange pentaquarks near the mass 1.5~GeV, initially reported by ZEUS using HERAI data \cite{Chekanov:2004kn}, was not confirmed using HERAII runs \cite{Abramowicz:2016xln}. 
The statistics for $K_S$ mesons and $K_S-p$ combinations for HERA II
was smaller than for HERAI, although some improvements in proton identification at HERA II were introduced using the microvertex detector, which could led to a higher purity for identified protons. 
Searches by the H1 collaboration \cite{Aktas:2006ic} using a similar statistics for $K_S-p$ combinations as in ZEUS did not confirm the ZEUS observation from HERAI. 
The results for strange pentaquarks from fixed-target experiments are typically mixed, i.e. had either positive or negative signals \cite{PhysRevD.98.030001}.
Therefore, the question of the existence of strange pentaquarks  is still open and represents an active area of
studies. The observation of lightest pentaquarks with strange quarks can be a significant discovery that can clarify the dynamics of multiquarks states. 

As for the strange pentaquarks,
a light charm pentaquark can shed  light on the dynamics of pentaquarks states. An evidence \cite{Aktas:2004qf} for  light charmed pentaquarks at around 3.1~GeV, first reported by H1, was not conformed by  ZEUS data \cite{Chekanov:2004qm}.

The main lesson for these searches of light pentaquarks in $ep$ collider experiments  is that evidences at  a 3~$\sigma$ level should always be confronted with the studies  using  alternative detector designs (ZEUS vs H1). 
Although the attempts to exclude the HERAI observations using  HERAII data is pointing to
the absence of the light pentaquarks, a firm conclusion can only be made using a statistics that is significantly larger than that achieved at the HERA collider. The EIC experiment has a significant potential to contribute to the searches for light pentaquarks.

\section{Forward physics and BFKL}
\begin{figure}[h!]
  \centering
  \includegraphics[width=0.75\textwidth]{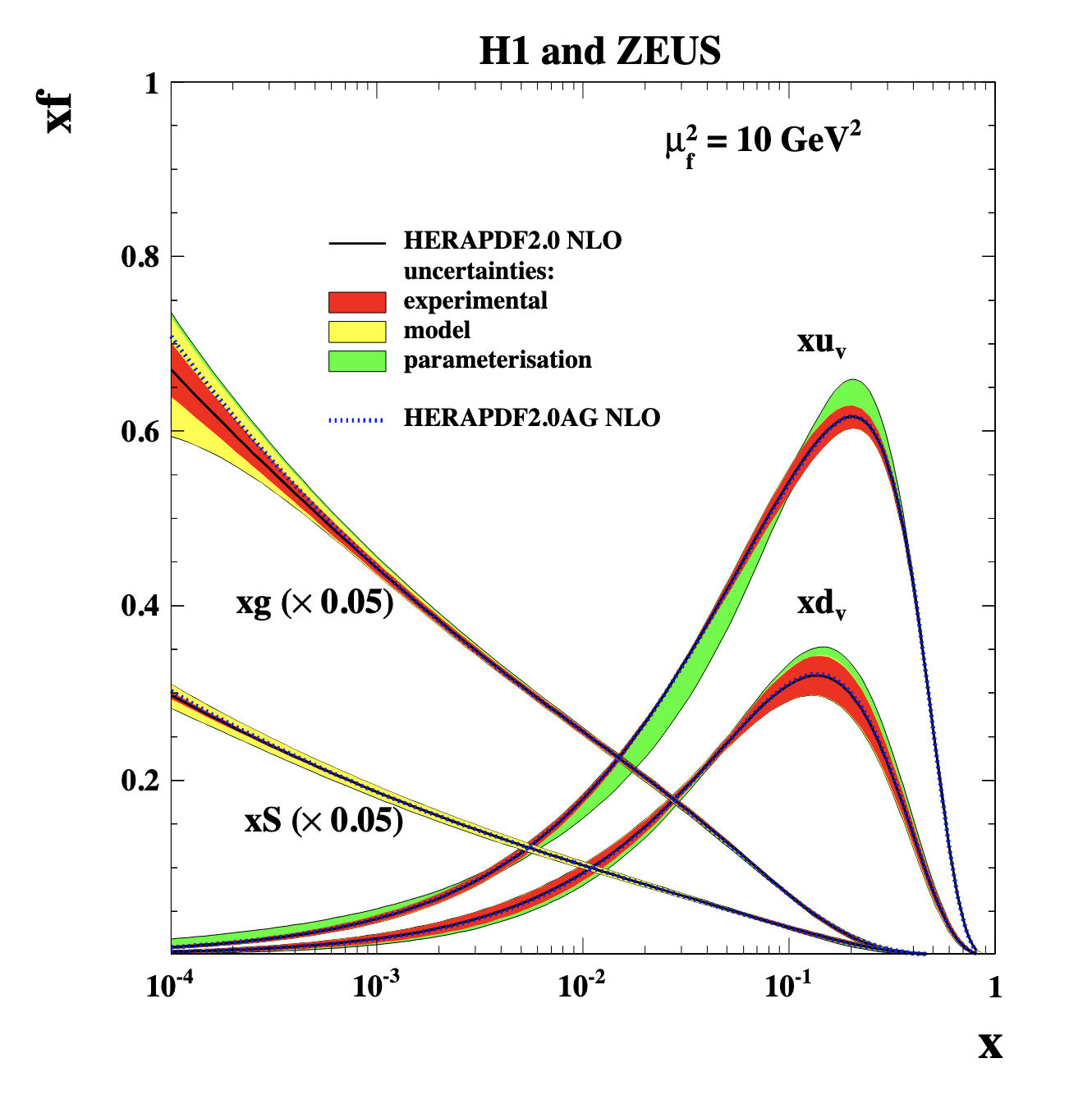}
  \caption{HERA H1 and ZEUS combined PDFs.}
  \label{xdec}
\end{figure}

One of the most significant findings of the experiments in DIS at HERA (ZEUS, H1) is the steep rise of the gluon density in the proton as the parton momentum fraction, $x$, decreases (see Fig.~\ref{xdec}) \cite{Klein_2008}.

Before HERA, there were many PDFs that exhibited the full range of the (then unknown) gluon density from flat to steeply rising at low $x$.  One consequence of a steeply rising gluon density is that at some point, gluons will completely fill the proton, which, if unchecked reaches a limit known as the unitarity bound, expressed as: 
\begin{equation}
    \label{unitarity}
    \alpha_s (x,Q^2) xg(x,Q^2)/Q^2 = \pi R^2.
\end{equation}
There must be some mechanism to stop or slow the gluon rise in order not to violate the unitarity limit as $x$ decreases.  As the gluon density saturates the proton, the unitarity limit can hold if gluons interact with each other, reducing the overall gluon density.  One particularly interesting idea of how this might manifest itself is for saturation effects to take place initially in small regions of the proton - so-called Hot Spots as described by A. H. Mueller in \cite{Mueller1991}.  No evidence was found for these regions of extreme QCD effects at HERA, but the use of large A nuclear targets at the EIC present an opportunity to revisit the ``hot spot'' saturation condition again.  

\begin{figure}[h!]
  \centering
  \includegraphics[width=0.75\textwidth]{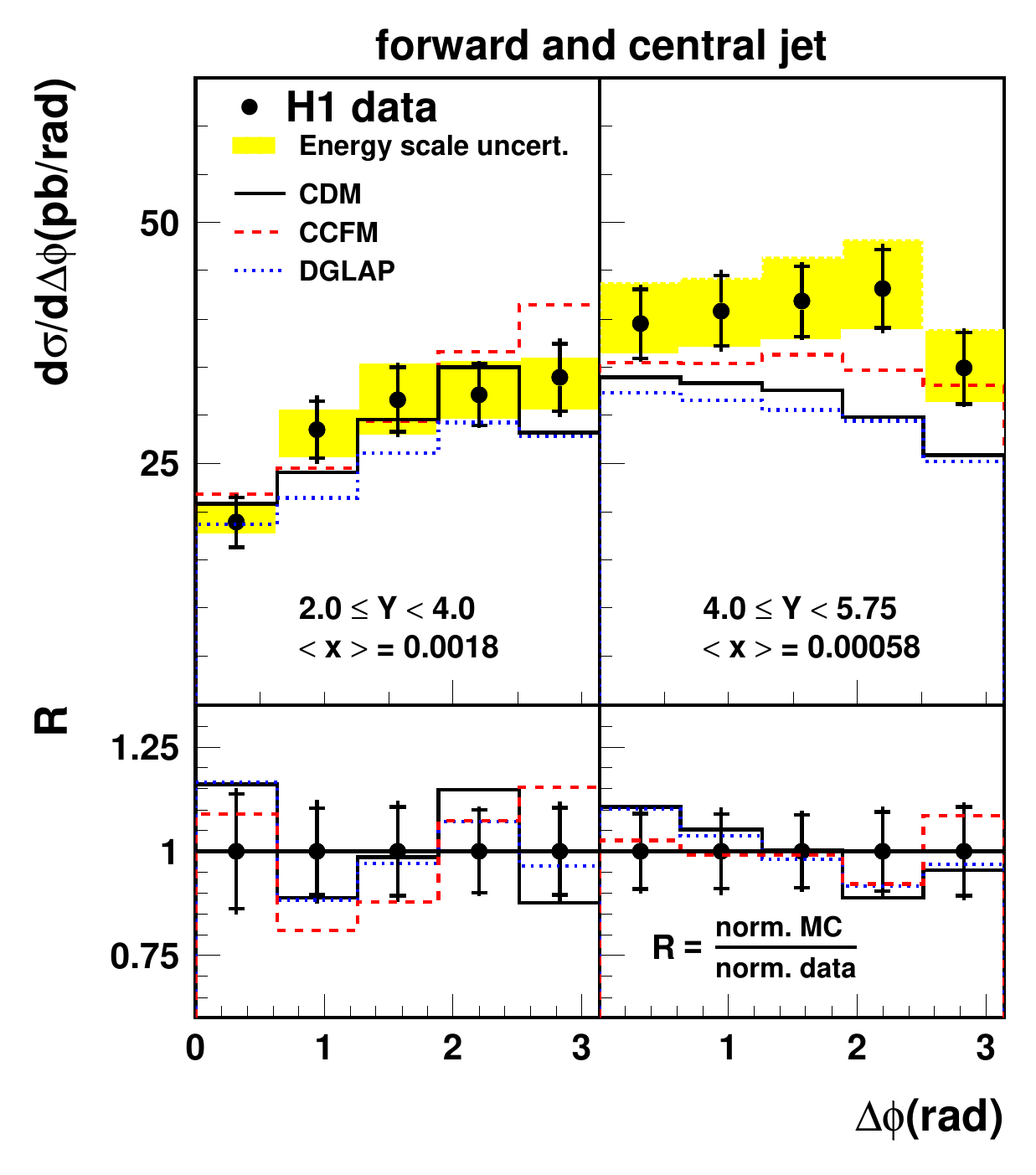}
  \caption{Forward and central jet cross sections as a function of the  azimuthal angle difference $\Delta \phi$
between the most forward jet and the scattered positron
in two intervals of the rapidity distance (reproduced from \cite{2012H1}).
}
\label{forwjH1}
\end{figure}

Recently, some aspects of the BFKL evolution have been discussed in \cite{jung2021aspects}. 
Jet measurements at HERA  show that the cross sections 
increase steeply towards small $x$, but the
predictions obtained from DGLAP parton shower simulations fall below the measurements.
Some experimental limitations for forward jets measurements at HERA 
are illustrated in Fig.~\ref{forwjH1} \cite{2012H1}.  The figure shows the prediction
of the three MC models.
The cross sections are measured in two intervals of rapidity distance ($Y$),
$2.0 \le Y < 4.0$  and $4.0 \le Y \le 5.75$.
The model with the key CCFM 
shows the CASCADE Monte Carlo generator  that 
implement the Ciafaloni-Catani-Fiorani-Marchesini (CCFM)
evolution that unifies the DGLAP and BFKL approaches.
 shows a better agreement with the data compared to  DGLAP based simulations
and the Colour Dipole Model (CDM) (see the corresponding references in \cite{2012H1}).

On the experimental side, it is clear that the forward-jet  
measurements suffer from the lack of statistics
and a significant ($4\%$) hadronic energy scale uncertainty that gives rise to the dominant
uncertainty of $7\%$ to $12 \%$ for the measured cross sections.
The EIC experiment can make a unique contribution to  such measurements 
if jet energy scale uncertainties will be reduced to a percent level.

\newpage
\section{Summary}

In conclusion, the EIC experiment offers unique opportunities to extend many physics measurements
performed in the previous circular colliders. 
This paper provides a snapshot of
several not-well understood physics effects encountered in such HEP experiments in the past, 
that can be verified and further studied at the EIC with a high precision. We believe that innovations in the accelerator 
and detector designs of the EIC, together with 
substantial theoretical progress in soft and hard QCD, could have a significant impact on understanding
of particle physics in the next two decades.

\section*{Acknowledgments}
The submitted manuscript has been created by UChicago Argonne, LLC, Operator of Argonne National Laboratory (“Argonne”). Argonne, a U.S. 
Department of Energy Office of Science laboratory, is operated under Contract No. DE-AC02-06CH11357. The U.S. Government retains for itself, 
and others acting on its behalf, a paid-up nonexclusive, irrevocable worldwide license in said article to reproduce, prepare derivative works, 
distribute copies to the public, and perform publicly and display publicly, by or on behalf of the Government.
The Department of Energy will provide public access to these results of federally sponsored research in accordance with the 
DOE Public Access Plan. \url{http://energy.gov/downloads/doe-public-access-plan}. Argonne National Laboratory’s work was 
funded by the U.S. Department of Energy, Office of High Energy Physics under contract DE-AC02-06CH11357.

\newpage
\section*{References}
\bibliographystyle{elsarticle-num}
\def\bibname{\Large\bf References}
\bibliography{references}

\end{document}